\documentclass[usenatbib]{mn2e}
\usepackage{graphicx}
\usepackage{aas_macros}
\usepackage{amsfonts}
\usepackage{bm}  
\usepackage{times}
\begin{document}


\def\beqnarray{\begin{eqnarray}}
\def\eeqnarray{\end{eqnarray}}
\def\norml{{\bf{\hat{l}}}}
\def\norms{{\bf{\hat{s}}}}
\def\vecv{{\bf{v}}}
\def\deg{\hbox{$\null^\circ$}}

\def\kpc{\,{\rm kpc}}
\def\Gyr{\,{\rm Gyr}}
\def\Myr{\,{\rm Myr}}
\def\kms{\,{\rm km}\,{\rm s}^{-1}}

\def\ud{\mathrm{d}}

\title[Are Complex A and the Orphan Stream related?]
{Are Complex A and the Orphan Stream related?}

\author[S.~Jin \& D.~Lynden-Bell]{Shoko Jin\thanks{e-mail: shoko@ast.cam.ac.uk} 
\& D. Lynden-Bell\\
Institute of Astronomy, University of Cambridge, Madingley Road, Cambridge, CB3 
0HA, U.~K.\\}
\date{}

\maketitle

\begin{abstract}
We consider the possibility that the Galactic neutral hydrogen stream
Complex A and the stellar Orphan stream are related, and use this
hypothesis to determine possible distances to Complex A and the Orphan
stream, and line-of-sight velocities for the latter.  The method
presented uses our current knowledge of the projected positions of the
streams, as well as line-of-sight velocities for Complex A, and we
show that a solution exists in which the two streams share the same
orbit.  If Complex A and the Orphan stream are on this orbit, our
calculations suggest the Orphan stream to be at an average distance of
$9\kpc$, with heliocentric radial velocities of approximately
$-95\kms$.  Complex A would be ahead of the Orphan stream in the same
wrap of the orbit, with an average distance of $4\kpc$, which is
consistent with the distance constraints determined through
interstellar absorption line techniques.

\end{abstract}

\begin{keywords}
Galaxy: structure --- Galaxy: halo --- ISM: clouds, kinematics and dynamics --- methods: analytical
\end{keywords}

\section{Introduction}
\label{sec:intro}

High velocity clouds (HVCs) have been enigmatic since they were first
discovered by \cite{HVCdiscovery}.  They are defined as neutral
hydrogen clouds which have anomalous velocities compared to what would
be expected through a simple model of differential Galactic rotation.
Most HVCs do not harbour stars and due to their general location away
from the Galactic disk, determination of distances to them through
interstellar absorption line techniques
\citep[e.g.][]{1997ARA&A..35..217W} is challenging.  The technique
has, however, been applied successfully to a handful of HVCs and a
distance `bracket' for the HVC can be obtained if both lower and upper
distance limits can be deduced.  The general lack of firm distance
determinations has, however, continued to make it difficult to
constrain theories for their origin.

Complex A is a system of high-velocity neutral hydrogen clouds and
resides in the northern Galactic hemisphere.  It has a distance
bracket of $\sim4.0--10\kpc$, determined through the detection and
non-detection of interstellar absorption lines at the velocities of
the clouds \citep[][]{vanWoerden1999Nature,1996ApJ...473..834W}.

The Orphan stream is a stellar stream discovered in the Sloan Digital
Sky Survey \cite[SDSS,][]{2000AJ....120.1579Y} Data Release 5 by
\cite{2007ApJ...658..337B}, and is located between Galactic coordinates
$l\sim 200\deg - 255\deg$ and $b\sim 52\deg - 48\deg$.  If the line of
Complex A is traced towards the stellar stream, the
two streams appear to be tantalisingly close to overlapping.

\cite{2007MNRAS.375.1171F} investigated the possibility of the dwarf
galaxy Ursa Major II (UMa II) being the progenitor of the Orphan
stream.  Their simulations of the orbit of the disrupting satellite
are able to reproduce the observational data to within the errors. In
their model, the Orphan stream is trailing behind UMa II with Complex
A being ahead of the dwarf galaxy by a further wrap of the orbit.  

As more stellar and gaseous streams continue to be found in the
Galaxy, it becomes increasingly interesting to investigate whether
associations between such streams can be found.  In this paper, we
consider the possibility that Complex A and the Orphan stream are
related, but without including UMa II in our analysis.  Instead, we
derive possible orbits for Complex A using its observational data and
show that it is possible for the two streams to be on the same wrap of
a single orbit.  This analysis then also tests the plausibility that
the streams are physically related.  Under this hypothesis, it is
possible to predict distances and line-of-sight velocities for the
Orphan stream and thereby test the validity of this scenario when
radial velocity data become available in the near future (Wilkinson
2007, priv. comm.).

\section{Method}
\label{sec:method}

In order to model the orbit of Complex A, we must have some realistic
initial conditions.  We assume Complex A to be a stream, in the sense
that it is moving along itself, and determine most of these initial
conditions from the observational data themselves.  A mid-stream point
is chosen as the `initial condition' (IC) point, for which we need the
sky-projected position, a heliocentric distance $d$ and the full
velocity vector in order for it to provide the initial conditions of
the orbit for Complex A.  The line-of-sight unit vector $\norml$ can
be determined since the Galactic longitudes and latitudes $(l,b)$ are
known.  Here and hereafter hats denote unit vectors.  The distance $d$
to the IC point is treated as a free parameter, with the aim of
obtaining a mutual orbit for Complex A and the Orphan stream.  The
line-of-sight velocities\footnote{This term will be used when
  referring to data from the modelled orbits.  To transform this to
  heliocentric radial velocities, we subtract the component of the
  Sun's velocity $[(9,232,7)\kms]$ along the line of sight to the
  position on the stream.  The distance from the Sun to the Galactic
  centre is taken to be $8.5\kpc$.} after correction to the Galactic
Standard of Rest, ${\bf{v}.\bf{\hat{l}}} = v_l$, for Complex A are
known \citep[e.g][]{1988A&AS...75..191H,1968BAN....20...33H}. In order
to obtain the full space velocity for the IC point, we must first
determine its velocity along the apparent stream (i.e. transverse to
the line-of-sight vector) as well.  We denote this component of
velocity in the plane of the sky as $v_s$, and the angle along the
stream, measured from one end of the stream, as $\chi$.  The unit
vector along the apparent direction of motion of the stream is given
by $\norms = \ud\norml/\ud\chi$ and $\ud\chi/\ud t = \vecv.\norms/d =
v_s/d$ from the usual equation for the transverse velocity.  Now
$\ud\vecv/\ud\chi$ is the acceleration $\ud\vecv/\ud t=\nabla\psi$
divided by $\ud\chi/\ud t$.  We then have the rate of change of
line-of-sight velocity along the stream:

\beqnarray
\label{eqn:radvelrun}
\frac{\ud v_l}{\ud\chi} \,=\, \vecv.\frac{\ud\norml}{\ud\chi} + \frac{\ud\vecv}{\ud\chi}.\norml
\,=\, v_s + (\norml.\nabla\psi)\frac{d}{v_s}\,,
\eeqnarray
where $\psi$ is the Galactic potential. Let $K(\norml) = \ud
v_l/\ud\chi$; we then have the following solutions to the quadratic
equation for $v_s$:

\beqnarray
\label{eqn:vs}
v_s = \frac{1}{2} \left( K(\norml) \pm \sqrt{K(\norml)^2 -
  4(\norml.\nabla\psi)d}\right)\,.  
\eeqnarray 
There are two possible solutions for $v_s$ and the two must be
distinguished through comparing the model orbit with the observational
data.  For the Galactic potential, we use a three-component Galactic
model described by \cite{1990ApJ...348..485P}, which uses the
Miyamoto-Nagai model for the disk and spheroid
\citep{1975PASJ...27..533M} and a near-logarithmic potential for the
halo.  In equation (\ref{eqn:vs}), both $\nabla\psi$ and $d$ naturally
depend on the assumed distance to the IC point.  We therefore treat
$d$ as a free parameter and compute both forward and backward
stretches of the orbit for different values of $d$.  For the distances
involved, our solutions are insensitive to the precise form of the
potential adopted, and we find, for example, that altering Paczynksi's
disk scale length $a_2$ by $\sim\pm35\%$ has a negligible effect on
the results obtained.

At any given position along Complex A, we can determine all components
of the velocity vector using one of the solutions for $v_s$ and the
observationally determined value of $v_l$.  The two solutions provide
orbits in opposite senses.  Both are physically valid but we find that
one can be ruled out on comparison of the computed orbit with the
observational data. 

\begin{figure}
\includegraphics[width=0.45\textwidth]{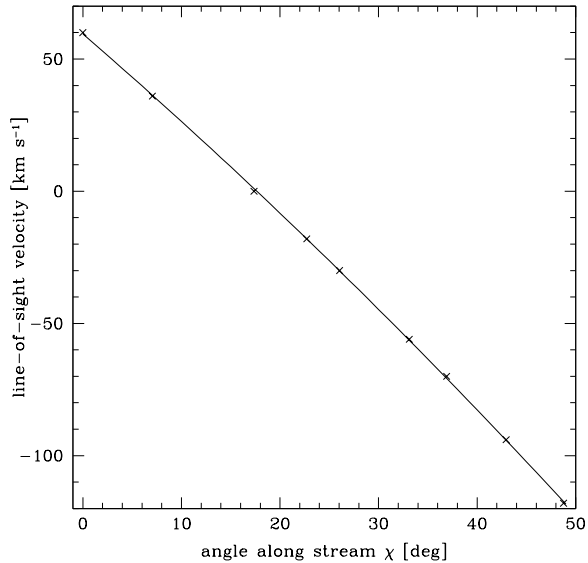}
\caption{Run of line-of-sight velocity along Complex A, shown as a
  function of angle along the stream.  $\chi = 0$ corresponds to the
  low-latitude end of the stream. The smoothed points for Complex A
  (crosses) have been fitted by a second order polynomial (solid line)
  and fit the data to $\pm7\kms$.
\label{fig:runradvel}}
\end{figure}

In order to calculate the run of radial velocity along the stream, it
is necessary to have a smoothed stream.  Given the morphology of
Complex A, we use the longitude and latitude combined, $(l+b)$, to
provide us with a single variable as a measure of position along the
stream.  Plotting $v_l$ and $l$ as a function of $(l+b)$ using data
from \cite{2004hvc..conf..195V} and \cite{1974IAUS...60..599D} and
taking a line of best fit through each provides us with a means of
extracting a smoothed stream.  The run of radial velocity along the
smoothed stream is shown in Figure \ref{fig:runradvel}, where we have
fitted a second order polynomial to the points.  The origin (zero point)
has been arbitrarily chosen to be the lower latitude end of the
stream.  The analytic fit allows us to determine $\ud v_l/\ud\chi$ and
hence the transverse\footnote{Transverse to the line of sight.}
velocity component along the smoothed stream.

Complex A fans out at higher latitudes and it is not possible to
constrain the orbit through points of this stream alone.  Our working
hypothesis is that the Orphan stream and Complex A share the same
orbit.  By computing possible orbits for Complex A at different
distances, we can determine whether it is possible for the stellar and
gaseous streams to share the same orbit.  If an orbit lies along both
streams, then that will constrain the distance to Complex A, and hence
the distance to the Orphan stream under this scenario.  We can
calculate the heliocentric radial velocities that the Orphan stream
stars would be expected to have, and these may then act as
`predictions' to test our model, once stellar radial velocity
measurements become available.

\section{Results}
\label{sec:results}

The two solutions for the tangential velocity $v_s$ in equation
(\ref{eqn:vs}) provide us with two possible orbits in the opposite
sense.  By comparing the observational dependence of $v_l$ on Galactic
coordinates and what the model predicts, we find that the orbit must
be in the same sense as that of Galactic rotation in order for 
our model to match the data for Complex A along its full length.

\begin{figure}
\includegraphics[width=0.45\textwidth]{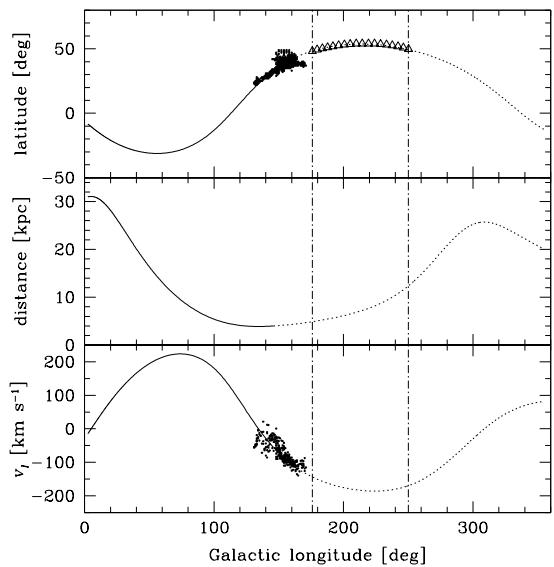}
\caption{Top: best fit orbit for Complex A (solid and dotted lines),
  overlaid with observationally determined positions for Complex A
  (dots) and the Orphan stream (open triangles). The solid line
  denotes the forward orbit from the initial condition point and the
  dotted line the backward orbit.  Initial conditions are $(l,b) =
  (146,33)\deg$, $d=4\kpc$, $v_l = -56\kms$ and $v_s =
  -266\kms$. Middle: heliocentric distances for the best fit orbit.
  Bottom: line-of-sight velocities for the same orbit, overlaid with
  observational data for Complex A (dots).  The dot-dash line shows
  the range covered by the Orphan stream.
\label{fig:fit}}
\end{figure}

We computed orbits with different IC points, varying the distance $d$
over a range of $1-40\kpc$.  Our best fit results are shown in Figure
\ref{fig:fit}, where the initial conditions at the position $(l,b) =
(146,33)\deg$ are given by $v_l = -56\kms$, $d = 4\kpc$ and
$\ud v_l/\ud\chi = -3.76 \kms/\rm{deg}$.  The solid curve shows the
forward-integrated orbit from the IC point and the backward
integration is shown by the dotted curve.  The value of $v_s$ at this
point is $-266\kms$, where the minus sign indicates that the direction
which we took to be increasing $\chi$ is opposite the actual direction
of motion of the stream. The orbital time covered in Figure
\ref{fig:fit} lasts in total for $470\Myr$.  The position of Complex A
\citep{1988A&AS...75..191H} is denoted by dots and the Orphan
stream illustrated by open triangles.  The Orphan stream is a straight
line going through (RA, dec) = $(162,0)\deg$ at an angle of
$(90-21.0375)\deg$ (Belokurov 2007, priv. comm.) and the positions
denoted in Figure \ref{fig:fit} were determined for 18 points between
RA of $145\deg$ and $162\deg$.

\section{Summary and Conclusions}
\label{sec:conclusions}

In this paper we have shown that by using the observational data of
one stream, it may be possible to constrain distances to more than
that stream.  The calculation uses the observational radial velocity
and positional data of the first stream together with the positional
data for the second.  Under an assumed Galactic potential, only the
distance to the first stream is treated as a free parameter.  Although
it is not possible to determine the distance to Complex A alone using
this method, the inclusion of a possible associated stream -- the
Orphan stream -- into the procedure introduces the means to calculate
distances to both streams and to predict radial velocities for the
second stream, under the hypothesis that they are related through the
same orbit.

Our results depend on the cores of the clouds of Complex A moving
ballistically.  The referee asked us to consider the possible effects
of ram pressure in altering the orbit of the gaseous stream, an effect
that would not be an issue for the stellar stream.  This would be
important in any crossing of the Galactic disk.  We estimate that the
clouds could travel for $5\kpc$ through a stationary medium of
particle density $\sim10^{-3}\rm{cm}^{-3}$ with a velocity change of
$\sim20\%$ if they were to sweep up all of the gas in its way.  The
effect would be much smaller if the halo gas and the HVCs are rotating
in the same direction or if we have overestimated the halo gas
density.  The effect of ram pressure is maximised in the above
calculation but it should be noted that this result is only an
order-of-magnitude estimate.

Our best fit results, assuming that Complex A and the Orphan stream
lie on the same orbit in a Galactic potential defined in Section
\ref{sec:method}, predict them to be at average distances of $4\kpc$
and $9\kpc$ respectively.  In this scenario, the two streams would be
in the same wrap of the orbit, with the direction of motion coinciding
with the direction of Galactic rotation and with Complex A ahead of
the Orphan stream.  The average heliocentric radial velocity of the
Orphan stream would be approximately $-95\kms$ with the velocity along
the stream ranging from $-155\kms$ on the side of Complex A to
$-30\kms$ at the other end.  \cite{2007MNRAS.375.1171F} show in their
Table 1 a summary of the distances and heliocentric velocities deduced
by \cite{2007ApJ...658..337B}.  We find that our distance predictions
increase in the opposite direction to theirs and that the values do
not agree to within their errors.  Two of their fields currently have
velocity estimates of $-35\kms$ and $105\kms$, but these are very
uncertain.

Vidrih et al. (2007, in prep.) have recently obtained radial velocity
data for the Orphan stream stars in regions that correspond to fields
2 and 3 in Figure 1 of \cite{2007ApJ...658..337B}, and should be able
to determine heliocentric radial velocities for stars in the region
between $(l,b)$ of $(215,54)\deg$ and $(242,51)\deg$.  As an example
of a location least contaminated with stars of the Sagittarius stream,
we predict the heliocentric radial velocity at $(l,b)$ of
$(230,53)\deg$ to be $-80\pm3\kms$. The indication of the error has
been determined from an uncertainty measure in the initial condition
distance of $\pm2\kpc$.  These predictions will be tested when Vidrih
et al. have reduced their data.

Many gaseous and stellar streams are known and continue to be found in
the Galactic halo.  If some of these seemingly disconnected streams
are in fact associated with each other, then it would be interesting
to have a method which attempts to find those associations.

\subsection*{Acknowledgements}

The authors thank V. Belokurov and M. Fellhauer for helpful
discussions on the Orphan stream, B. Wakker for the kind use of data
for Complex A and the anonymous referee for helpful comments.  SJ
acknowledges financial support from PPARC.

\end{document}